\pdfoutput=1
 \documentclass[aps,prl,reprint,twocolumn,showpacs,superscriptaddress]{revtex4-1}  

 \usepackage{auncial}
\usepackage{scrextend}
\usepackage{relsize}
\usepackage{amsmath}
\usepackage{amssymb}
\usepackage{epsfig}
\usepackage{graphicx}
\usepackage{hyperref}
\usepackage{dcolumn}   
\usepackage{tabu}
\usepackage{boldline}
\usepackage{slashed}
\usepackage{multirow}
\usepackage{color}
\usepackage[normal]{subfigure}
\usepackage{rotating}
\usepackage[margin=0.9in,a4paper]{geometry}
\usepackage[table]{xcolor}
\usepackage{enumitem}
\usepackage[utf8x]{inputenc}
\usepackage{colortbl}
\usepackage{array,multirow}
\definecolor{nicered}{rgb}{0.7,0.1,0.1}
\definecolor{nicegreen}{rgb}{0.1,0.5,0.1}
\definecolor{red}{rgb}{1.0, 0, 0}
\def\mX{{\mathcal{X}}}

\def\eq#1{{Eq.~(\ref{#1})}}
\def\eqs#1#2{{Eqs.~(\ref{#1})--(\ref{#2})}}

\def\Tr{\mbox{Tr}\,}
\def\Min{\mbox{Mnr}\,}
\def\diag{\mbox{diag}\,}
\def\det{\mbox{det}\,}
\def\arg{\mbox{arg}\,}

\def\gsim{\raise0.3ex\hbox{$\;>$\kern-0.75em\raise-1.1ex\hbox{$\sim\;$}}}
\def\lsim{\raise0.3ex\hbox{$\;<$\kern-0.75em\raise-1.1ex\hbox{$\sim\;$}}}

\def\mb[#1]{\mathbf{#1}}

\renewcommand{\bar}{\overline}

\definecolor{LightCyan}{rgb}{0.88,1,1}
\definecolor{piggypink}{rgb}{0.99, 0.87, 0.9}
\definecolor{applegreen}{rgb}{0.55, 0.71, 0.0}
\definecolor{darkpastelgreen}{rgb}{0.01, 0.75, 0.24}
\definecolor{green-yellow}{rgb}{0.68, 1.0, 0.18}

\newcommand{\beq}{\begin{equation}}
\newcommand{\eeq}{\end{equation}}
\newcommand{\beqa}{\begin{eqnarray}}
\newcommand{\eeqa}{\end{eqnarray}}
\newcommand{\Sec}[1]{ \medskip \noindent {\sl \bfseries #1}}

\newcommand{\eqn}[1]{eq.~(\ref{#1})}

\begin{document}


\title{Accidental Peccei-Quinn symmetry protected to arbitrary order}

\author{Luca Di Luzio}
\email{luca.di-luzio@durham.ac.uk}
\affiliation{\normalsize \it 
Institute for Particle Physics Phenomenology, Department of Physics, Durham University, DH1 3LE, United Kingdom}
\author{Enrico Nardi}
\email{enrico.nardi@lnf.infn.it}
\affiliation{\normalsize\it INFN, Laboratori Nazionali di Frascati, C.P.~13, 100044 Frascati, Italy}
\author{Lorenzo Ubaldi}
\email{ubaldi.physics@gmail.com}
\affiliation{\normalsize\it Raymond and Beverly Sackler School of Physics and Astronomy, Tel-Aviv University, Tel-Aviv 69978, Israel}

\begin{abstract}
  \noindent
  A $SU(N)_L\times SU(N)_R$ gauge theory for a scalar multiplet $Y$
  transforming in the bi-fundamental representation $(N,\bar N)$
  preserves, for $N>4$, an accidental $U(1)$ symmetry firstly broken
  at operator dimension $N$.  Two configurations are possible for the
  vacuum expectation value of $Y$, which correspond to the (maximal)
  little groups $\mathcal{H}_s=SU(N)_{L+R}$ and
  $\mathcal{H}_h=SU(N-1)_L\times SU(N-1)_R\times U(1)_{L+R}$.  In the first
  case the accidental $U(1)$ gets also broken, yielding a pseudo
  Nambu-Goldstone boson with mass suppression controlled by $N$, while
  in the second case a global $U(1)$ remains unbroken.  The strong CP
  problem is solved by coupling $Y$ 
  to new fermions carrying color.  The first case allows for a
  Peccei-Quinn solution with $U(1)_{PQ}$ protected up to order $N$ by
  the gauge symmetry.  In the second case $U(1)$ can get broken by
  condensates of the new strong dynamics, resulting in a composite
  axion.  By coupling $Y$ to fermions carrying only weak isospin,
  models for axion-like particles can be constructed.
\end{abstract}


 \maketitle

 \Sec{Introduction.}  
 In the past decades, a plethora of experimental results has firmly
 established QCD as the correct description of strong interaction
 phenomena in particle physics. However, together with a deep
 understanding of many fundamental issues, this beautiful theory also
 brings in one theoretical conundrum. The QCD gauge sector depends on
 two dimensionless parameters whose value is not predicted by the
 theory, but must be determined experimentally. The first one,
 $\alpha_s$, determines the strength of the QCD interactions. Its
 measured value is a natural one for a dimensionless quantity
 (roughly speaking is of order unity, although the precise number depends
 on the energy scale). 
 The second one, $\theta$, gives the amount of CP violation in strong
 interactions.  Theory only dictates that $\theta$, which is an
 angular variable, must fall within the interval $[0,2\pi)$, and
 also in this case it would be natural to expect $\theta\sim
 O(1)$. Instead, experimental limits on the neutron electric dipole
 moment 
 yield the upper bound $|\theta|<10^{-10}$, a value that is regarded
 as highly unnatural. This theoretical nuisance bears the name of
 ``the strong CP problem''.  QCD, however, would recover its
 naturalness if, for some reason, $\theta=0$. 
\hfill\break
An elegant mechanism to guarantee the vanishing of $\theta$ was
proposed in 1977 by Peccei and Quinn
(PQ)~\cite{Peccei:1977ur,Peccei:1977hh}.  It relies on a $U(1)_{PQ}$
global symmetry, anomalous with respect to QCD, broken spontaneously
by the vacuum expectation value (VEV) of a Standard Model (SM) singlet
scalar field at a scale $v_a \gg 100\,$GeV, and broken explicitly by
non-perturbative QCD effects at a scale $\Lambda_{QCD}\sim
100\,\text{MeV}$.  Spontaneous breaking (SB) of a global $U(1)$
symmetry gives a massless Nambu-Goldstone boson (NGB). However, due to
the presence of a relatively tiny explicit breaking, the NGB of
$U(1)_{PQ}$ is not exactly massless: it is a pseudo NGB, commonly
referred to as the {\it axion}~\cite{Weinberg:1977ma,Wilczek:1977pj}.
To account for $|\theta|< 10^{-10} $ any other source of explicit
$U(1)_{PQ}$ breaking besides QCD must either be absent or adequately
suppressed.  This is difficult to achieve, especially considering that
$U(1)_{PQ}$, being anomalous, is not even a real symmetry.  Thus,
effective operators not respecting $U(1)_{PQ}$ are expected to arise.
Even if suppressed by the Planck scale $m_P = 1.2 \cdot
10^{19}\,\text{GeV}$, unless their dimension is larger than $d\sim 10$
they would unavoidably give 
$|\theta| > 10^{-10}$~\cite{Georgi:1981pu,Kamionkowski:1992mf,Holman:1992us,Barr:1992qq}.

In this Letter we propose a mechanism that, on the basis of first
principles, can protect $U(1)_{PQ}$ to arbitrary accuracy.  A scalar
multiplet $Y$ is assigned to the bi-fundamental representation
$(N,\bar N)$ of the gauge group $G_{LR}^{(N)}=SU(N)_L\times
SU(N)_R$. For $N>4$ an accidental $U(1)$ symmetry corresponding to
phase transformations of $Y$ is enforced at the classical level, and
it only gets broken at $d=N$ by the determinant interaction $\det(Y)$.
When the scalar gauge theory is coupled to fermions carrying color,
$U(1)$ acquires a QCD anomaly.  SB of $G_{LR}^{(N)}$ via a VEV of $Y$
can proceed via two patterns. In the first case $U(1)$ also undergoes
SB, acquiring all the features of a PQ symmetry: $\theta$ becomes a
fundamental dynamical field with a periodic potential that drives its
value to zero.  In the second case a global $U(1)'$ remains
perturbatively unbroken. However, condensates of the new
strong gauge dynamics can break it, giving rise to a composite
axion. In both cases a solution to the strong CP problem is obtained.

\Sec{Accidental $U(1)$ in $G_{LR}^{(N)}$ scalar gauge theory.}
For $N>4$, $G_{LR}^{(N)}$ gauge invariance restricts the
renormalizable potential for the scalar multiplet $Y$
to the following simple form:
\beq
\label{V0}
V_0 = \lambda \left[T- 
v_a^2/2\right]^2 + \lambda_A\,A \,, 
\eeq
where $v_a$ is a constant with the dimension of a
mass, $T= \Tr[YY^\dagger]$ and $A= \Tr[\Min(YY^\dagger,2)]$, with
$\Tr[\Min(M,k)]$ denoting the trace of the matrix of the minors of
order $k$ of $M$~\footnote{The reader might be more familiar with the
  double trace $\Tr[YY^\dagger YY^\dagger]=T^2-2A$ as a second
  invariant.}. We require $v_a^2>0$ to trigger SB
and $|\lambda_A| < \frac{2N}{N-1}\lambda$ to ensure that the potential
is bounded from below.  The matrix $Y^c$ of constant
background values of $Y(x)$ can be written in its singular value
decomposition as:
\begin{equation}
\label{eq:Yc}
\frac{\sqrt{2}}{v_a}\,Y^c = \mathcal{U}_L \,\hat Y \, \mathcal{U}_R^\dagger
= U_L  \left(
\hat\Phi \hat Y\right)  U_R^\dagger \ \rightarrow\  \hat \Phi\hat Y, 
\end{equation}
where $\hat Y = \diag(y_1,\,y_2,\dots ,y_N)$ is diagonal with real
non-negative entries normalized such that $\sum_iy_i^2=1$, $\mathcal{U}_{L,R}$
are unitary matrices, $U_{L,R} $ are special unitary
($\det(U_{L,R})=+1$), $\hat\Phi$ 
is a diagonal matrix of phases 
such that 
$\log\det(\hat\Phi)=\log\det(\mathcal{U}_L\mathcal{U}_R^\dagger) =i\,\arg\det(Y^c)\equiv
i\delta^c $.  
$\delta^c$, being an angular variable, ranges in the interval
$[0,2\pi)$, and physics must be invariant under the redefiniton
$\delta^c \to \delta^c + 2 n \pi$ ($n=1,2\dots\,.$). The last
(diagonal) form in \eq{eq:Yc} is obtained, without loss of generality,
via a rigid $G_{LR}^{(N)}$ rotation.  We will leave understood that
$Y^c$ is always written in this basis as $Y^c = (v_a/\sqrt{2}) \,
\hat\Phi\,\hat Y$.

The vacuum configurations that minimize $V_0$ 
are easily found~\cite{Nardi:2011st}: $T$ is blind to specific $\hat
Y$ configurations (this is because it carries a $SO(2 N^2)$ accidental
symmetry much larger than the gauge symmetry that allows to rotate
between different configurations). Minimization of the first term then
just fixes the ``length'' $\ell(\hat Y) = v_a^{-1}\sqrt{2\langle
  T\rangle}=1$.
The extrema of $\langle A\rangle \propto \sum_{i<j} y_i^2y_j^2$
instead depend on the structure of $\hat Y$. We have two
possibilities: {\rm(i)} for $\lambda_A<0$, $\langle A\rangle$ is
maximized at the symmetric point $y_i^2 = 1/N, \ \forall i$; {\rm
  (ii)} for $\lambda_A>0$ the minimum occurs when $\langle
A\rangle=0$, that is 
when all entries in $\hat Y$, but one, vanish. In summary, the
configurations that extremize $V_0$ are:
\begin{eqnarray} 
\nonumber
\label{Ysh}
{\rm(i)} \quad Y^c_s =\frac{v_a}{\sqrt{2}} \, \hat\Phi\, \hat Y_s \, , 
&\quad& \hat Y_s=\frac{1}{\sqrt{N}}\diag\left(1,1,\dots ,1\right) \, , \\
{\rm(ii)} \quad Y^c_h =  \frac{v_a}{\sqrt{2}} \, \hat\Phi\, \hat Y_h \, ,
&\quad& \hat Y_h=\diag\left(0,\dots , 0,1\right) \, .
\end{eqnarray}
The corresponding little groups are the two maximal subgroups of
$G_{LR}^{(N)}$: $\mathcal{H}_s= SU(N)_{L+R}$ for $Y^c_s$ and
$\mathcal{H}_h= G^{(N-1)}_{LR}\times U(1)_{L+R}$ for $Y^c_h$, where
the Abelian generator corresponds to the diagonal combination of the
LR Cartan generators of $G_{LR}^{(N)}$ proportional 
to $\lambda^{L,R}_{N^2-1} = \diag(1,1,\dots , 1-N)_{L,R}$. 
It is important to stress that $\mathcal{H}_{s,h}$ cannot get broken
further by any type of perturbative effects~\cite{Georgi:1977hm} or,
equivalently, that neither the vanishing entries in $Y_h^c$ can be
perturbatively lifted, nor the strict equality of the entries in
$Y_s^c$ can be perturbatively spoiled.  Some bibliographic remarks
are in order: the minima of the potential for the case of global
$G_{LR}^{(3)}$ (namely the SM quark flavor symmetry)
were studied in~\cite{Nardi:2011st} (and with the assumption of a real
$\det(Y)$ previously in \cite{Conversi:1970fua}).  The possibility of
raising perturbatively the vanishing entries in $Y_h$ to generate the
SM fermion mass hierarchy was addressed in~\cite{Espinosa:2012uu}. It
was found that, in agreement with the Georgi-Pais (GP)
theorem~\cite{Georgi:1977hm}, minimization of the effective potential
results in the same little groups $\mathcal{H}_{s,h}$. Only by
introducing additional reducible scalar representations a more
thorough breaking, yielding $y_{i \neq j} \neq y_j$, can be
obtained~\cite{Espinosa:2012uu,Fong:2013dnk}.

The tree level potential $V_0$ in \eqn{V0} has an accidental $U(1)$
rephasing symmetry $Y\to e^{i (\alpha/N)} Y $ (under which $\delta^c
\to \delta^c+\alpha$) so that the full symmetry of the classical
Lagrangian is in fact $G_{LR}^{(N)} \times U(1)$.  The first minimum
$Y^c_s$ breaks this symmetry and yields a NGB, which, in first
approximation, remains massless.
However, accidental symmetries are generally not respected by
operators of higher dimensions. Here it is the requirement of
local gauge invariance that dictates at which order these operators
can arise.  A fundamental set of higher order operators can be
constructed by considering the characteristic polynomial
$\mathcal{P}(\xi)$ of the matrix $YY^\dagger$:
\beqa
\mathcal{P}(\xi) = \det\left( \xi I - YY^\dagger\right) 
= 
\sum_{n=0}^N (-1)^n C_n\, \xi^{N-n},\quad 
\label{Cn}
\eeqa
where $I$ is the identity matrix, and $C_{n}=
\Tr[\Min(YY^\dagger,n)]$, with $C_0=1$, $C_1=T$, $C_2=A$, \dots, $C_N
= \det[YY^\dagger]\equiv |\mathcal{D}|^2 $.  The solutions of
$\mathcal{P}(\xi)=0$ are the eigenvalues of $YY^\dagger$ and, being
the eigenvalues invariant under $G^{(N)}_{LR}$, so are the
coefficients $C_{n}$. They correspond to invariant combinations of
components of $Y$ of dimension $d=2n$~\cite{Espinosa:2012uu}.
The determinant $\mathcal{D} = \det Y$ is another invariant, since
under $G^{(N)}_{LR}$, $\mathcal{D} \to \det (V_L\,Y\,V^\dagger_R)=\det
Y$~\footnote{All other invariants can be expressed in terms of the set
  $ \{ C_1(x),C_2(x),\dots , C_{N-1}(x),\mathcal{D} \} $,
  see~\cite{Espinosa:2012uu}.}.  However, while all $C_n$'s respect
the $U(1)$ accidental symmetry, under $Y\to e^{i(\alpha/N)}Y$,
$\mathcal{D}(x)\to e^{i\alpha}\mathcal{D}(x)$.  Thus, $U(1)$ gets
firstly broken at $d=N$ by:
\begin{equation}
\label{VD}
V_{D} =\frac{k\, \mathcal{D} + k^*\, \mathcal{D}^*}
{m_P^{N-4}}
=\frac{2 \kappa D }{m_P^{N-4}}  \cos[\varphi + \delta(x)]\,, 
\end{equation}
%
where $\kappa$ and $\varphi$ are the modulus and argument of the
coupling $k$, $D=|\mathcal{D}|$, $\delta(x) = \arg \mathcal{D}(x)$,
and the $m_P^{N-4}$ suppression stems from the assumption that $V_D$
is only generated by gravity effects.  Let us see which is the fate of
the NGB of  case~(i).  The minimum of $V_D$ is obtained for
$\langle \delta(x)\rangle\equiv \delta^c = \pi-\varphi$, and the
minimum of $V_0$ is lowered by the amount:
\begin{equation}
  \label{eq:DeltaV}
\Delta V = v_a^4 \frac{2\kappa}{(2N)^{N/2}}
\left(\frac{v_a}{m_P}\right)^{N-4}.   
\end{equation}
Thus, in the breaking $G_{LR}^{(N)} \times U(1) \to \mathcal{H}_s$,
$N^2-1$ of the initial $2(N^2-1) + 1$ generators are left unbroken,
$N^2 -1$ are spontaneously broken with the corresponding NGB eaten by
gauge bosons that acquire masses $O(v_a)$, while the NGB of the global
$U(1)$ acquires a tiny mass $O(\sqrt{\Delta V}/v_a)$ because of the
explicit breaking \eq{VD}. In case~(ii) instead, a global $U(1)'$
generated by $\lambda^{L+R}_{N^2-1} + (N-1) I$ is preserved by
$Y^c_h$, so that at the renormalizable level
$G_{LR}^{(N)} \times U(1) \to \mathcal{H}_h\times U(1)'$.  Although
the higher order operator \eq{VD} breaks $U(1)'$ in interactions,
since in the ground state $\langle D\rangle =0$, there is no SB and no
 NGB arises. This case, in which the symmetry of the
effective Lagrangian is smaller than the symmetry of the
renormalizable Lagrangian, but the resulting little group is the same,
provides a non trivial test of the GP theorem~\cite{Georgi:1977hm}.

\Sec{Solutions to the strong CP problem.} 
Solutions to the strong CP problems can be implemented by introducing
fermions carrying color. 
Two different types of solutions are
possible, depending on which vacuum is selected in the breaking of
$G_{LR}^{(N)}$.
Let us proceed by steps by first introducing four fermion multiplets
transforming under $G_{LR}^{(N)}$ as: $Q_L \sim (N,1)$, $Q_R \sim
(1,N)$, $\Psi_L \sim (\bar N,1)$ $\Psi_R \sim (1,\bar N)$.  Since they
can be combined into real representations of $SU(N)_{L,R}$
($Q_{L,R}\oplus\Psi_{L,R}$), there are no gauge anomalies.  Gauge
symmetry allows for Yukawa couplings of the form $\bar Q_L Y Q_R +
\bar\Psi_L Y^\dagger \Psi_R + \text{H.c.}$. They preserve the $Y$
rephasing symmetry if the fermions are transformed chirally with
$U(1)$ charges satisfying: $\mX_{Q_L} - \mX_{Q_R} = \mX_Y$ and
$\mX_{\Psi_L} -\mX_{\Psi_R} = - \mX_Y $.  The opposite sign of the two
charge differences (a consequence of the requirement of gauge
anomalies cancellation) ensures the absence of $U(1)$-$SU(N)_{L,R}$
mixed anomalies~\footnote{This point is important for the PQ solution:
  it ensures that the VEV of the axion will not end up canceling the
  $SU(N)_{L,R}$ $\theta$-terms rather than $\theta_{QCD}$.}.  Let us
now triplicate the fermion content, and assign $Q_{L,R}$ to the
fundamental representation of color, while $\Psi_{L,R}^a$ ($a=1,2,3$)
remain color singlets.  Since there is no compensating cancellation of
the $Q_{L,R}$ contribution, a $U(1)$-QCD anomaly arises, and
$U(1)$ acquires all the features required for a PQ symmetry.

\medskip

\noindent
{\it (i) Solution with a fundamental axion.}\  
We choose a basis in which the SM quark masses are real while
$\theta_{QCD}\neq 0$.  Without loss of generality the $\Psi_{L,R}$
couplings can be taken flavor-diagonal, so that the Yukawa terms
can be written as:
\begin{equation}
  \label{eq:Lm}
e^{i(\eta_0/N)} h_0\, \bar Q_L Y Q_R +
   e^{i(\eta_a/N)} h_a \bar \Psi_L^a Y^\dagger \Psi_R^a \,,
\end{equation}
where $h_0,h_a$ ($a=1,2,3)$ are four real non-negative parameters.  If
$\lambda_A<0$, the minimum $Y_s^c$ is selected, and all the fermions
become massive, with degenerate masses within each $SU(3)_c$ and
$SU(N)_{L+R}$ multiplet.  We first show that the fermion masses
stemming from \eqn{eq:Lm} can be brought into real form without
inducing mixed $G^{(N)}_{LR}$ anomalies. 
After SB of the gauge symmetry, $\arg\det(M_Q) = \eta_0+\delta^c$ and
$\arg\det(M_a) = \eta_a-\delta^c$, that is there are four independent
phases that we wish to cancel (four conditions).  We can perform four
chiral rotations of the fermion multiplets respectively with phases
$\alpha_0,\alpha_a$, subject to a fifth condition $\sum_{a=1}^3
\alpha_a =3\alpha_0$ which avoids mixed anomalies with the $SU(N)$
gauge groups. The phase of $Y$ can be also redefined (this changes the
argument of the cosine in \eqn{VD} by the addition of a constant term
$\varphi \to \tilde\varphi$).  All the complex phases can thus be
canceled. However, the chiral rotation of $Q_{L,R}$ is anomalous with
respect to $SU(3)_c$ of color, and  another source of explicit $U(1)$
breaking is then introduced.
Including it, the relevant potential for $\delta(x)$ acquires the form:
\begin{equation}
  \label{eq:Vd}
  V_\delta = \Delta V \cos[\tilde\varphi+\delta(x)]
  - f_\pi^2 m^2_\pi \cos [\delta(x)] \, ,
\end{equation}
where $\tilde\varphi$ is a generic constant unrelated with
$\theta_{QCD}$, and we have redefined $\delta(x)+\theta_{QCD} \to
\delta(x)$ so that the anomalous coupling to the gluons can be 
written as $\frac{\alpha_s}{4 \pi} \delta \,G \tilde G$.  
%
%
%
Note that when the angular variable $\delta(x)$ is varied in the
interval $[0,2\pi)$, there is a unique minimum of the potential.
Namely, independently of $N$, the number of domain
walls~\cite{Sikivie:1982qv} is always one.
%
From \eq{eq:Vd} we see that if $\kappa,\tilde\varphi=O(1)$, as it is
natural to assume, $\delta^c <10^{-10}$ can be ensured only if the
(gravitationally induced) explicit breaking satisfies $\Delta
V/(f^2_\pi m^2_\pi) \lsim 10^{-10}$.  For the phenomenologically
preferred interval $10^9\, \text{GeV} \lsim v_a/N \lsim 10^{12}\,
\text{GeV}$ this condition can be fulfilled with $9\leq N\leq 13$~\footnote{For $N>10$, QCD asymptotic freedom is lost 
beyond the heavy quarks' threshold. However, Landau poles in $\alpha_s$ remain trans-Planckian for $N \lesssim 26$.}.

\begin{figure}[t!]
\centering
\includegraphics[angle=0,width=7.5cm,height=6.8cm]{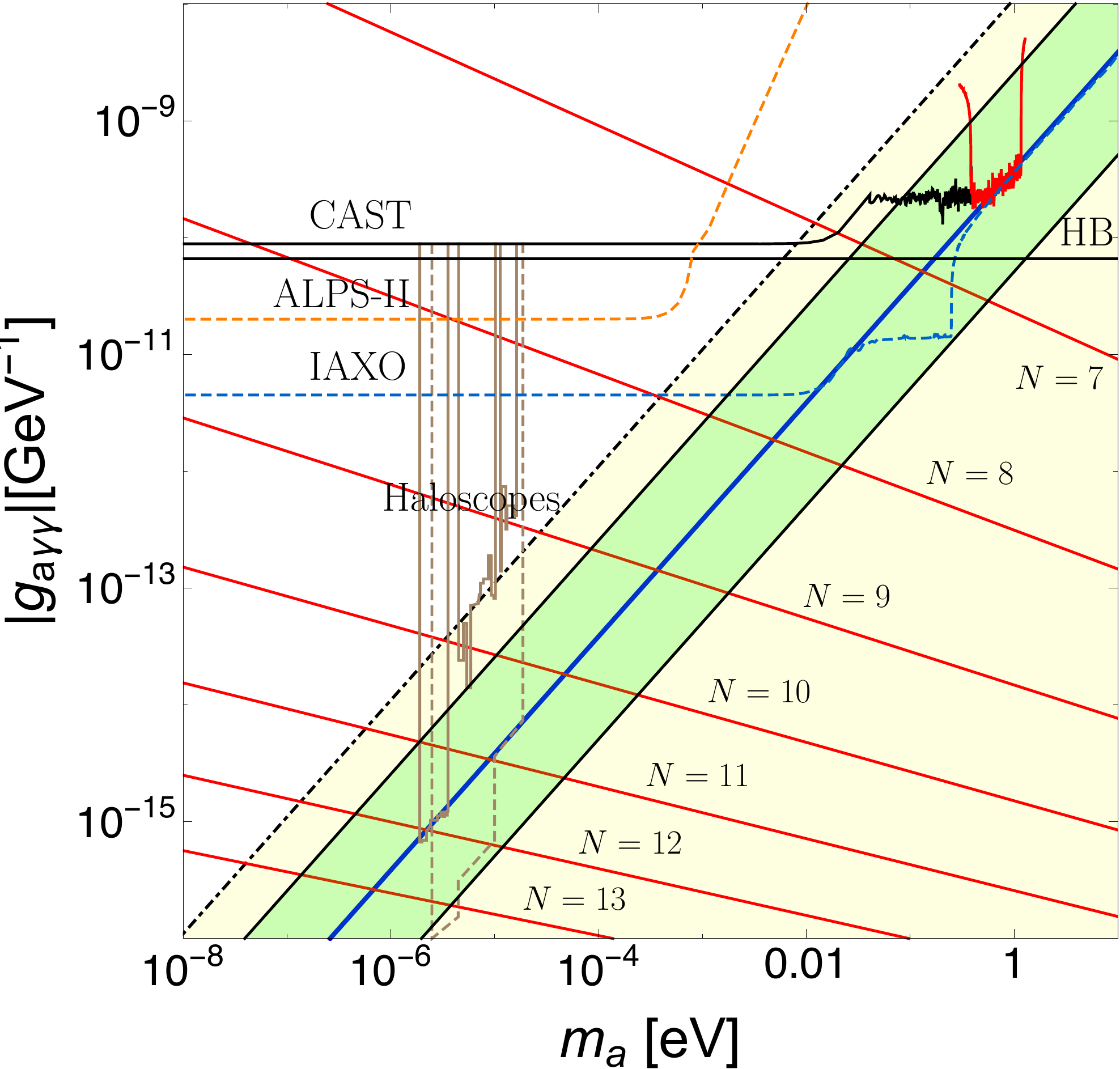}
\caption{\label{fig:ALP} 
ALP-photon coupling versus mass for different 
choices of the gauge group $G_{LR}^{(N)}$.
The green band and the yellow area represent two regions for 
phenomenologically preferred axion models~\cite{DiLuzio:2016sbl,inpreparation}.
The axion coupling in \eq{eq:agg} corresponds to the thick blue line. 
}
\end{figure}

%
Let us now proceed to identify the
fundamental axion field and some of its properties.  
In the ``unitary'' gauge in
which the rigid $G_{LR}^{(N)}$ rotation yielding 
$Y^c \sim \hat\Phi\hat Y $ in \eq{eq:Yc} 
is promoted to a local one, we can write $Y(x)=\hat\Phi(x) \hat Y(x)$ 
with 
\begin{equation} 
\label{Phix}
\hat\Phi(x) = \diag(e^{i \hat\gamma_1(x)}\!, \dots , e^{i
    \hat\gamma_N(x)}); \;  \ 
             \hat\gamma_i = \frac{\sqrt{2 N}}{v_a} \gamma_i. 
\end{equation}
The linear combinations of the $N$ ``orbital'' modes $\hat\gamma_i$
corresponding to $N-1$ non-Abelian broken generators and
to the accidental $U(1)$ are:
\begin{equation}
  \label{eq:fields}
a_a(x) 
= 2\,\Tr\left[\vec\gamma(x) \cdot T_a\right] \, , 
\end{equation}
where $\vec\gamma = (\gamma_1,\dots ,\gamma_N)$ and, for $a=1,\dots ,N-1$, 
$T_a$ are the $SU(N)$ Cartan generators with normalization 
$\Tr[T_a]^2 = 1/2$, while $T_0 =(1/\sqrt{2N})I$.  Then, the canonically
normalized axion field is:
\begin{equation}
  \label{eq:axfield}
 a_0(x) 
=  2\,\Tr\left[\vec\gamma \cdot T_0\right]
 =  \frac{v_a}{N} \delta(x)\,. 
\end{equation}
%
Note that since the periodicity of $\delta(x)$ is $2\pi$, the
periodicity of the axion field is $a_0\to a_0 + \frac{2\pi}{N}v_a$.
One might wonder whether, contrary to what stated below \eq{eq:Vd},
there are $N$ domain walls corresponding to the $N$ minima $\langle
a_0 \rangle + \frac {2\pi n}{N} v_a$, ($n=0,\dots ,N-1$).  This is not
so: all these minima are in fact gauge equivalent, in the sense that
the $Z_N$ center of $SU(N)_{L+R}$ has precisely as elements $ \exp({i
  2\pi n/N})\cdot I$, so that the cyclic values of $a_0/v_a$ can be
all connected via gauge transformations.
Neglecting the subdominant gravitational contributions,  
the mass of the axion is $m_a = N (m_\pi f_\pi)/ v_a$, while the
strength of its coupling to the photon via the usual term 
$(1/4) g_{a\gamma\gamma} F_{\mu\nu} \tilde F^{\mu\nu}$ is:
\begin{equation}
  \label{eq:agg}
g_{a\gamma\gamma} =   -1.92\,  \frac{m_a}{\text{eV}}\,
\frac{2.0}{10^{10} \ \text{GeV}}\,. 
\end{equation}
which falls within the  axion 
window, see  Fig.~\ref{fig:ALP}.

\smallskip

\noindent
{\it (ii) Solution with a composite axion.}
If in \eqn{V0} $\lambda_A>0$, the minimum $Y_h^c$ provides mass for
just one fermion in each $N$-dimensional multiplet, $12(N-1)$ Weyl
fermions remain massless, and a global $U(1)'$ acting on these
massless fermions remains unbroken.  It is then tempting to try to
implement a massless-quark type of solution, where the topological
term is simply removed via a chiral $U(1)'$ rotation without other
consequences on the Lagrangian. As regards the massless $Q$'s and
$\Psi$'s, they would get confined into $F$-hadrons once
$G_{LR}^{(N-1)}$ enters the strongly coupled regime at a scale
$\Lambda_F\gg \Lambda_{QCD}$.  However, such a scenario is not viable
because matching the $U(1)^\prime$-QCD anomaly in the low energy
theory~\cite{tHooft:1980xss} requires that some composite fermions
carrying color remain massless, and these are not observed.  We are
then led to assume that $U(1)'$ gets spontaneously broken by some
color neutral condensate of $Q$ and $\Psi$. In this case the pseudo
NGB of the $U(1)'$ symmetry would correspond to a composite
axion~\cite{Kim:1984pt,Randall:1992ut,Redi:2016esr} with mass and
couplings suppressed as  $1/\Lambda_F$. Clearly, this second scenario is
more speculative, so that we will mainly focus on the first scenario.
%

\Sec{Phenomenology.}
In the first scenario with a fundamental axion, after SB the spectrum
consists of:
\begin{labeling}{\ $\bullet$} \itemsep -2pt     
\item[\ $\bullet$] $N^2-1$ gauge bosons with masses $O(v_a)$; 
\item[\ $\bullet$] $N$ quarks $Q$ and $3N$ SM singlet fermions
  $\Psi^a$, stable at the tree level, with $m_{Q,\Psi}\sim O(v_a/\sqrt{N})$;
\item[\ $\bullet$] $N^2-1$ massless gauge bosons $F$; 
\item[\ $\bullet$] One pseudo NGB (the axion).
\end{labeling}
\vspace{-3pt}
%
%
%
%
%
Tree level stability of the heavy $Q$'s and $\Psi$'s follows from the
fact that they do not carry weak isospin and hypercharge
$\mathcal{Y}$, and thus cannot decay into lighter SM fermions. Different
representations allowing for couplings between the $Q$'s and the SM
fermions would bring the problem of canceling
$\left[G_{LR}^{N}\right]^2\times U(1)_{\mathcal{Y}}$ gauge anomalies,
a complication that we prefer to avoid.  However, cosmologically
stable heavy relics, and in particular long lived strongly interacting
particles like the $Q$'s, represent serious issues in cosmology and
astrophysics (see~\cite{DiLuzio:2016sbl,inpreparation} for a recent
discussion and relevant references).
A simple way to avoid all phenomenological problems is to assume a
pre-inflationary scenario (PQ symmetry broken before inflation) and
$v_a > T_{\text{reheating}}$ so that, after inflation has wiped away
all the heavy states, they cannot be regenerated.  This holds, in
particular, for the heavy $Q$ which otherwise would be copiously
produced via QCD interactions.  As regards the massless gauge bosons
$F$,
their production after inflation could proceed via gravitational
effects, or via gluon-gluon scattering into a pair of $F$'s mediated
by
loops of heavy $Q$'s.  The first mode is suppressed by powers of $m_P$
and typically very inefficient.  The rate for the second process can
be estimated as $(\alpha_F\,\alpha_s)^2\, T^9/m^8_Q$, with $\alpha_F$
the coupling strength of the new gauge group.  This reaction remains
well out of equilibrium for all $T< m_Q$ 
and thus also this channel is typically too inefficient to produce $F$
in sizable amounts. All in all, we can conclude that in
pre-inflationary scenarios no dangerous heavy relics are left in
appreciably amounts.

Post-inflationary scenarios (PQ symmetry broken after inflation, and
$v_a < T_{\text{reheating}}$) yield a different picture.  During
reheating, all the heavy states can attain equilibrium
distributions.  At $T \lsim v_a$ the heavy gauge bosons with masses
$O(v_a)$ will readily decay into the lighter $Q,\Psi$ fermions. 
Below  $m_Q$, the unbroken $SU(N)_{L+R}$ corresponds to
an effective pure Yang-Mills theory with large $N$ 
that rapidly flows towards a confining regime at a scale $\Lambda_F\gg
\Lambda_{QCD}$.
Bound state mesons $\Pi_{Q\,(\Psi)} \sim Q\bar Q\, (\Psi\bar \Psi$)
singlets under $SU(N)_{L+R}$ form, and readily decay into lighter
gaugeballs $G \sim FF$ of mass $m_G\sim O(\Lambda_F)$ ($\Pi_{Q}$ must
be also color singlets).  These `gaugeballs' are easily disposed of:
they can decay into two gravitons with a rate $\Gamma_{grav} \sim
\Lambda_F^5/m_P^4$~\cite{Soni:2016gzf},
%
%
or into a pair of gluons via heavy quark loops with a rate
$\Gamma_{gg} \sim \Lambda^9_F/m_Q^8$.  If $m_Q/m_P \lsim \Lambda_F/
m_Q$, decays into gluons proceed faster than decays into gravitons.
%
%
%
However, visible decays are subject to severe constraints from big
bang nucleosynthesis and from several other observations (limits on
CMB spectral distortions, diffuse $\gamma$-rays background, etc.)
which, taken together, suggest $\tau_G \lsim 10^{-2}\,s.$ Rewriting
the lifetime as $\tau_G \sim 10^{-33}\,
\left(10^{9}\text{GeV}/m_Q\right) \left(m_Q/\Lambda_F\right)^9 s.$ we
see that safe lifetimes are obtained if the heavy quark mass and
confinement scale fulfill $m_Q/\Lambda_F \lsim 3\times 10^{3}$, a
condition that can be easily satisfied for $N \gsim 9$.  However,
there are other more dangerous heavy relics, like $\mathcal{M}_{ab}
\sim \Psi_a\bar{\Psi}_{b}$ with $(a\neq b)$ and, since at $\Lambda_F$
color is unconfined, `mongrel' mesons $\mathcal{M}_a \sim Q\Psi_a$
will also form.  $\mathcal{M}_{ab} $ decays are forbidden by
$\Psi$-flavor conservation, while decays of $\mathcal{M}_a$ are
forbidden also by color conservation.  The abundance of these states
is basically determined by free particle annihilation before
$F$-confinement, which always results in $\Omega_{\mathcal{M}} \gg
\Omega_{DM}$ unless their overall mass scale is brought down to values
not much larger than a few TeV. This requires tiny values for the
Yukawa couplings in~\eqn{eq:Lm} and an appropriately small initial
values for the $G_{LR}^{(N)}$ gauge couplings to ensure
$\Lambda_F\lsim $few TeV. All in all, the post-inflationary
scenario, if not ruled out, is certainly strongly disfavored with
respect to pre-inflationary scenarios.

\Sec{Axion like particles.}
With no attempt to solve the strong CP problem, models for axion like
particles (ALPs) can be constructed along the same lines.  Instead
than new quarks, let us introduce new colorless fermions
$\mathcal{T}$, doublets under weak-isospin and, for simplicity, with
zero hypercharge. SM singlet fermions $\Psi^a_{L,R}$ ($a=1,2$) are
also introduced to cancel gauge anomalies.
%
%
Now, in the breaking $G_{LR}^{(N)}\to \mathcal{H}_s$ the NGB of the
accidental $U(1)$ only receives mass from the gravity induced
determinant operator of $d=N$. However, here $N$ does not need to be
particularly large so that, compared to axions, much larger ALP masses
and photon couplings are possible. For the ALP mass we obtain:
\beq
\label{maALPs}
m_a = \frac{N \sqrt{2 \kappa}}{(2N)^{N/4}} 
\left(\frac{v_a}{m_P} \right)^{\frac{N-4}{2}} v_a \, ,
\eeq
and for the ALP-photon
coupling we have:
\begin{equation}
\label{agg}
g_{a\gamma\gamma} = \frac{\alpha}{2\pi}\frac{E}{v_a} \,, 
\end{equation}
with 
the electromagnetic anomaly coefficient:
\beq 
\label{eq:E}
E = 
2\, N\, \mX_{\mathcal{T}_L} \mathcal{Q}^2_{\pm} 
= \frac{N}{2}
\,,  
\eeq
where $\mathcal{Q}_\pm =\pm\frac{1}{2}$ are the electric charges of
the components of $\mathcal{T}_L$, and 
$\mX_{\mathcal{T}_R}=0$, $\mX_{\mathcal{T}_L}=\mX_Y=1$
the (assigned) PQ charges.
\eqs{maALPs}{eq:E} yield
\beq 
\label{ALPwindowformula}
g_{a\gamma\gamma} = 
B_N\, m_a^{-\frac{2}{N-2}} \, ,
\eeq 
with  
\beq 
\label{defCN}
B_N = 
\frac{N\,\alpha }{4\pi} \left( 
\frac{2 \kappa N^2}{(2N)^\frac{N}{2}m_P^{N-4}}
\right)^{\frac{1}{N-2}} \, .
\eeq
The ALP-photon coupling versus $m_a$ is plotted in Fig.~\ref{fig:ALP}
for different values of $N$.  The preferred regions for the
axion~\cite{DiLuzio:2016sbl,inpreparation} are also shown.

\Sec{Conclusions.} 
We have put forth a new realization of the PQ solution to the strong
CP problem. Our scenario might be loosely classified as a KSVZ type of
axion model~\cite{Kim:1979if,Shifman:1979if} since PQ charges are
carried only by non-SM particles.  A new gauge group
$G_{LR}^{(N)}=SU(N)_L\times SU(N)_R$ is postulated, new quarks
$Q_{L,R}$ are assigned to fundamental representations of $SU(N)_{L,R}$
while the PQ scalar, rather than being a single complex field, is a
matrix $Y$ transforming in the bi-fundamental representation of
$G_{LR}^{(N)}$.  A PQ symmetry arises accidentally, and remains
protected by the gauge symmetry from all types of explicit breaking up
to dimension~$N$, which is in principle arbitrary.  Within this same
construction, and depending on the gauge symmetry breaking pattern, a
different possibility where the axion is composite can be realized.
%

%

\section*{Acknowledgments}

E.N.~thanks G.~Villadoro for enlightening conversations.  We
acknowledge S.~Tulin for pointing out the relevant literature on
glueballs DM, and M.~Nardecchia for discussions.  
E.N.~is supported by 
%
%
the INFN ``Iniziativa Specifica'' TAsP.  L.D.L.~and E.N.~acknowledge
the organizers of the CERN-EPFL-Korea Theory Institute ``New Physics
at the Intensity Frontiers'', where part of this work was carried out,
for invitation and financial support.

 \bibliographystyle{apsrev4-1.bst}

%

\end{document}